\UseRawInputEncoding
 \documentclass[aps,prl,twocolumn,groupedaddress]{revtex4-1}
\usepackage{bm}
\usepackage[dvipdfmx]{graphicx} 
\usepackage{here}
\usepackage[usenames]{color}
\usepackage{amsmath,amsfonts,amssymb}
\usepackage{hyperref}
\usepackage{xcolor}
\usepackage{float}

\begin{document}
\title{
		Topological band theory of a generalized eigenvalue problem with Hermitian matrices: Symmetry-protected exceptional rings with emergent symmetry
		}

\author{Takuma Isobe$^1$}
\author{Tsuneya Yoshida$^{1,2}$}
\author{Yasuhiro Hatsugai$^{1,2}$}
\affiliation{
$^1$Graduate School of Pure and Applied Sciences, University of Tsukuba, 
Tsukuba, Ibaraki 305-8571, Japan\\
$^2$Department of Physics,
University of Tsukuba, Tsukuba, Ibaraki 305-8571, Japan
}

\date{\today}

\begin{abstract}
So far, topological band theory is discussed mainly for systems described by eigenvalue problems.                              
Here, we develop a topological band theory described by a generalized eigenvalue problem (GEVP).
Our analysis elucidates that non-Hermitian topological band structures may emerge for systems described by a GEVP with Hermitian matrices.
The above result is verified by analyzing a two-dimensional toy model where symmetry-protected exceptional rings (SPERs) emerge although the matrices involved are Hermitian.
Remarkably, these SPERs are protected by emergent symmetry, which is unique to the systems described by the GEVP.
Furthermore, these SPERs elucidate the origin of the characteristic dispersion of hyperbolic metamaterials which is observed in experiments.
\end{abstract}

\maketitle

\textit{Introduction.---}
After discovery of topological insulators, topological band structures have been studied as one of the central issues in condensed matter systems~\cite{C.L.Kane_E.J.Mele_PRL.2005,C.L.Kane_E.J.Mele_PRL.2005_Z2,L.Fu_C.L.Kane_PRL.2007,M.Z.Hasan_C.L.Kane_RevModPhys.2010,X.L.Qi_S.C.Zhang_RevModPhys.2011,Y.Ando_JPSJ_2013,B.A.Bernevevig_T.L.Huglhes_S.C.Zhang_Science_2006,M.Knig_Science_2007,L.Fu_C.L.Kane_PRB.2007,L.Fu_C.L.Kane_PRB.2006,D.J.Thouless_PRB.1983,Schnyder_PRB.2008,A.Y.Kitaev_AIP_Conf_2009,S.Ryu_A.P.Schnyder_A.Furusaki_New.J.Phys_2010,X.L.Qi_T.L.Hughes_S.C.Zhang_PRB_2008,A.M.Essen_J.E.Moore_D.Vanderbilt_PRL_2009}. 
In these systems, robust gapless modes emerge around the boundary due to topological properties in the bulk~\cite{Hatsugai_PRL93}.
In addition, the topological band theory is also applicable to semimetals which host robust band touching in the bulk~\cite{S.Murakami_IOP_2007,W.Xiang_PRB_2011,Yang_PRB_2011,A.Birkov_PRL_2011,Xu_PRL_2011,Kurebayashi_JPSJ_2014,N.Armitage_RevModPhys_2018,Koshino_PRB_2016}. 
For instance, the Chern number can be assigned to the gapless points in the bulk of Weyl semimetals, which elucidates novel transport properties~\cite{W.Xiang_PRB_2011,Yang_PRB_2011,A.Birkov_PRL_2011,Xu_PRL_2011}.
Mathematically speaking, these topological band structures are described by a standard eigenvalue problem with a Hermitian matrix.

Recently, the topological band theory has been extended to non-Hermitian systems~\cite{K.Esaki_PRB_2011,M.Sato_Progress_of_Theoretical_Science_2012,S.-D.Liang_PRA_2013,D.Leykam_PRL_2017,Xu_PRL(2017)_WeylEP,Shunyu_PRL(2018)_SkinEffect,Gong_class_PRX18,Yao_nHChern_PRL2019,Kawabata_nHclass_PRX19,Yoshida_nHFQH19,xiao_NatPhysics(2020)_Observsation-nHBBC,xiao_arXiv(2020)_observation-nBPTsymm,Dibyendu_PRB(2021)_nH-SSH} which are described by the standard eigenvalue problem with a non-Hermitian matrix. 
The platforms of the non-Hermitian topological band theory are extended to even beyond quantum systems such as photonic crystals with gain/loss~\cite{Bo_zhen_nature_2015_Ering,Takata_pSSH_PRL18,Ozawa_TopoPhoto_RMP19,Junpeg_PRL_2020_HMMs}, electric circuits~\cite{Hofmann_ExpRecipSkin_19,Helbig_ExpSkin_19,Hofmann_EleCirChern_PRL19,Yoshida_MSkinPRR20}, mechanical metamaterials~\cite{Yoshida_SPERs_mech19,Scheibner_nHmech_PRL2020} and so on. 
Remarkably, the non-Hermiticity of such systems enriches topological properties. 
For instance, non-Hermitian systems may host exceptional points where both for the real- and imaginary-parts of energy bands touch~\cite{H.Shen_PRL_2018,VKozii_nH_arXiv17,T.Yoshida_PRB_2018,A.A.Zyuzin_PRB_2018,Michishita_EP_PRB2020,Yoshida_nHReview_PTEP20}. 
It has also been elucidated that symmetry of non-Hermitian systems results in symmetry-protected exceptional rings (SPERs)~\cite{Bergholtz_PRB_2019,T.Yoshida_PRB_2019,Delplace_EP3_arXiv2021} and symmetry-protected exceptional surfaces (SPESs)~\cite{Okugawa_SPERs_PRB19,T.Yoshida_PRB_2019,Zhou_SPERs_Optica19,K.Kimura_PRB_2019} in two- and three- dimensions, respectively, although preserving the relevant symmetry requires fine-tuning.

Along with the above development of the topological band theory, recent studies have revealed that several systems are described by generalized eigenvalue problems (GEVPs)~\cite{Raghu_PhC_PRL2008,Raghu_PhC_PRA2008,Shindou_GEV_PRB2013}.  
The above progress of topological band theory implies the presence of novel topological phenomena unique to GEVPs. 
Unfortunately, however, most of the previous works focus on the case where the problem is reduced to the ordinary Hermitian systems. 
Thus, for systems described by GEVPs, a further development of a topological band theory remains a crucial issue to be addressed.

In this letter, we discuss the topological band theory for systems of a GEVP with Hermitian matrices. 
Our analysis elucidates that such systems may exhibit non-Hermitian topological phenomena protected by emergent symmetry. 
As an example, we demonstrate the emergence of SPERs for a system described by a GEVP with Hermitian matrices. 
Notably, no fine-tuning is necessary to realize these SPERs because they are protected by emergent symmetry.
This property is unique to systems described by the GEVPs.
Furthermore, the SPERs with emergent symmetry explain the origin of the hyperbolic dispersion of hyperbolic metamaterials (HMMs)~\cite{Smith_PRL_2003_HMMs,Smith_APL_2004_HMMs,Liu_OptExp(2008)_HMMs,Fang_PRB(2009)_HMMs,Poddubny_NatPhoto_2013_HMMs,Drachev_OptExp_2013_HMMs,Shekhar_2014_HMMs,Ferrari_PiQE_2015_HMMs,Guo_AIP_2020_HMMs,Noginov_APL(2009)_HMMs,Kanungo_APL(2010)_HMMs,Krishnamoorthy_Science(2012)_HMM,Starko_JOptSocAmB(2015)_HMMs,Kruk_NatComm_2016_HMMs,W.Ji_PhysRevMaterials(2020)_HMMs}.

So far, SPERs and SPESs have been reported for non-Hermitian systems described by standard eigenvalue problem.
We would like to stress, however, that SPERs and SPESs described by the GEVP with Hermitian matrices do not require fine-tuning in order to preserve the relevant symmetry, which is a striking difference from the ordinary SPERs.

\vskip\baselineskip

\textit{GEVP with Hermitian matrices.---} 
Here, we analyze a general theory of a GEVP with Hermitian matrices describing non-Hermitian topology.
Specifically, we elucidate that a complex band structure may emerge despite the Hermiticity of matrices.
We also show that generalized eigenenergies form pairs due to emergent symmetry.

Let us consider the band theory described by a GEVP, which is defined as~\cite{footnote6}
\begin{equation}\label{eq:GEVP}
H\psi=E S\psi.
\end{equation}
Here, $H$ and $S$ are Hermitian matrices, $E$ are generalized eigenvalues and $\psi$ are generalized eigenvectors~\cite{footnote7}.
Unless otherwise noted, generalized eigenvalues and generalized eigenvectors are simply referred to as eigenvalues and eigenvectors in this letter.

In the following, we show that when $H$ and $S$ are indefinite,
$E$ can be complex in spite of the Hermiticity of $H$ and $S$.
Without loss of generality, we assume that $S$ is diagonalized.
Here, matrix $S$ can be decomposed as
\begin{equation}
S=S^{\prime}\Sigma S^{\prime},
\end{equation}
with $S^{\prime}=\mathrm{diag}(\sqrt{|\beta_1|},\dots,\sqrt{|\beta_n|})$,
$\Sigma=\mathrm{diag}[\mathrm{sgn}(\beta_1),\dots,\mathrm{sgn}(\beta_n)]$, and
$\beta_i$ being eigenvalues of matrix $S$.
$\mathrm{sgn}(\beta_i)$ take the sign of $\beta_i$.
When $S$ is indefinite,
$\Sigma$ is not propotional to the identity matrix.
Here, we define $\phi$ and $\tilde{H}$ as $\phi=S^{\prime}\psi, \tilde{H}=S^{\prime -1}HS^{\prime -1}$.
Because of the relation
$\Sigma=\Sigma^{-1}$, Eq.~(\ref{eq:GEVP}) is rewritten as
\begin{equation}
H_{\Sigma}\phi=E \phi,
\end{equation}
with $H_{\Sigma}=\Sigma\tilde{H}$.
The matrix $H_{\Sigma}$ is non-Hermitian, and
eigenvalues $E$ are given by complex.
Noticing Hermiticity of $\tilde{H}$ and $\Sigma$, we can find emergent symmetry; $H_{\Sigma}$ satisfies the relation,
\begin{equation}
\Sigma^{-1}H_{\Sigma}\Sigma=H_{\Sigma}^{\dagger},
\end{equation}
which is known as pseudo-Hermiticity.
Therefore, eigenvalues $E$ are real or form complex conjugate pairs~\cite{Mostafazadeh_pHvsPT1_2002,Mostafazadeh_pHvsPT2_2002,Mostafazadeh_pHvsPT3_2002} (for details, see Sec.~I of Supplemental Material~\cite{supple}).
We note that the indefiniteness of $H$ and $S$ is essential for the complex band structure; either $H$ or $S$ is definite, eigenvalues are always real~\cite{footnote1}.

In the above, we have shown that system described by the GEVP with Hermitian matrices may exhibit a complex band structure. 
The emergence of the complex band structure can be understood by mapping the GEVP to the standard eigenvalue problem~\cite{footnote9}. We stress, however, that such a mapping yields pseudo-Hermiticity as emergent symmetry, which is a significant difference from ordinary systems described by the standard eigenvalue problem with a non-Hermitian matrix.

\vskip\baselineskip

\textit{SPERs in a two-band model.---}
We have seen that systems, described by the GEVP with Hermitian matrices show complex eigenvalues.
In the following, we specifically analyze a two-band model to demonstrate
SPERs, a unique topological band touching of non-Hermitian topological bands.

\begin{figure}[t]
   \includegraphics[width=8cm]{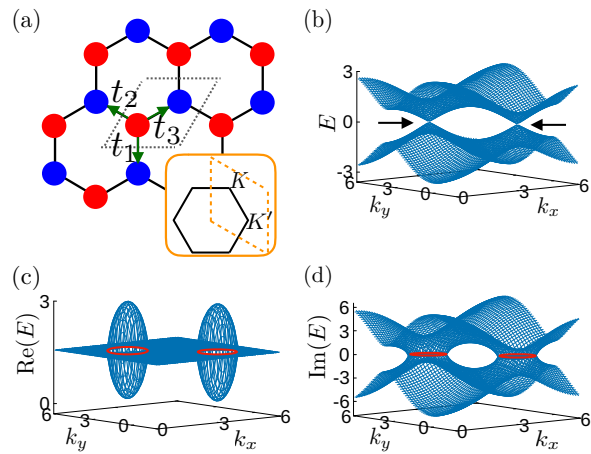}
 \caption{
  (a): Sketch of the honeycomb lattice model and its Brillouin zone. The gray (orange) dashed lines describe the unitcell (first Brillouin zone).
	$\bm{t}_1$,$\bm{t}_2$, and $\bm{t}_3$ are vectors connecting neighboring site.
  These vectors are defined as $\bm{t}_1=(0,-1)$, $\bm{t}_2=(-\sqrt{3},1)/2$, and $\bm{t}_3=(\sqrt{3},1)/2$.
  (b): The dispersion relation of overlaped honeycomb lattice model with $m_L=0$, $m_R=0$.
  $v$ is fixed at $1$.
	Dirac points at $K$ and $K'$ points are denoted by arrows.
 (c)[(d)]: The real [imaginary] part of the dispersion relation of overlaped honeycomb lattice model with $m_L=0.3$, $m_R=1.1$. 
 SPERs (red ring) based on the GEVP emerges around the $K$ and $K'$ points.
 }
 \label{fig:2x2model}
\end{figure}

Here, we consider a two-band model of the honeycomb lattice [see Fig.~\ref{fig:2x2model}(a)].
We note that such a model described by $2\times 2$ Hermitian matrices is the minimal model to investigate the SPERs.
The GEVP of the model is written as
\begin{equation}\label{eq:2x2model}
\begin{pmatrix}
m_L&vf_{\bm{k}}\\
vf_{\bm{k}}^* & -m_L
\end{pmatrix}
\psi=E
\begin{pmatrix}
1+m_R&0\\
0 & 1-m_R
\end{pmatrix}
\psi,
\end{equation}
with the hopping $v$.
Here, $m_L$ and $m_R$ are real~\cite{footnote8}.
The eigenvalues (eigenvectors) are denoted by $E$ ($\psi$).
$f_{\bm{k}}$ is defined as $f_{\bm{k}}=(1+e^{i\bm{k}\cdot\bm{t_1}}+e^{i\bm{k}\cdot\bm{t_2}})$
with $\bm{t}_1$, $\bm{t}_2$, and $\bm{t}_3$ being vectors connecting neighboring sites [see Fig. \ref{fig:2x2model}(a)].
Here, we take the matrix in the left-hand (right-hand) side of Eq.~(\ref{eq:2x2model}) as $H$ ($S$).

Now let us discuss the band structure.
Eigenvalues of this model are described by
\begin{align}
E=E_0\pm\sqrt{M^2-|g_{\bm{k}}|^2},
\end{align}
with $E_0 = [m_L/(1+m_R)- m_L/(1-m_R)]/2$,
$M=[m_L/(1+m_R)- m_L/(1-m_R)$]/2, and
$g_{\bm{k}}=(vf_{\bm{k}})/\sqrt{1-m_R^2}$.
Details of analysis are provided in Sec.~II of Supplemental Material~\cite{supple}.
In the following, we discuss the band structure for three cases; (i)$ |m_R|<1$, (ii) $|m_R|>1$, and (iii) $|m_R|=1$.
We note that in cases (i), (ii), and (iii), matrix $S$ is definite, indefinite, and not invertible, respectively.

We start with case (i) where the model shows Dirac cones at $K$ and $K'$ points [see Fig.~\ref{fig:2x2model}(b) which is obtained for $m_L=0$ and $m_R=0$].
Here, the band structure has a mass gap with non-zero $m_L$ and $m_R$ for $|m_R|<1$.


Next, we discuss the case (ii).
Figures~\ref{fig:2x2model}(c)~and~\ref{fig:2x2model}(d) show the band structure for $m_L=0.3$ and $m_R=1.1$.
In these figures, we can find the SPERs where both of the real- and the imaginary-parts show band touching around $K$ and $K'$ points. 
The topological characterization of SPERs can be done by computing the zero-th Chern number $N_{0\mathrm{Ch}}$, which is the number of negative eigenvalues of Hermitian matrix $\Sigma (H_{\Sigma}-E_{\mathrm{ref}})$ with $E_{\mathrm{ref}}=1.57$, $\Sigma=\mathrm{diag}(1,-1)$, and 
$
H_{\Sigma}=
\left(
\begin{array}{cc}
{m_L}/{|1+m_R|} & {vf_{\bm{k}}}/{\sqrt{|1-m_R^2|}}  \\
{-vf^*_{\bm{k}}}/{\sqrt{|1-m_R^2|}} & {m_L}/{|1-m_R|}
\end{array}
\right)
$.
Specifically, inside (outside) of the ring, the zero-th Chern number takes $N_{0\mathrm{Ch}}=$2 (1) which topologically protects the non-Hermitian band touching. 
We note that the above SPERs are protected by the emergent symmetry; they are robust against perturbations preserving Hermiticity of $H$ and $S$, and translational symmetry.
For case (iii) (i.e., for $m_R=1$), $S$ is not invertible, and one of the eigenvalues diverges.

In the above, we have demonstrated the emergence of SPERs for the system described by a GEVP with $2\times2$ Hermitian matrices.
Intriguingly, the topology of SPERs is protected by emergent symmetry.
Here, we stress that for the emergence of SPERs, the indefiniteness of both matrices $H$ and $S$ are essential.
Although the overlap matrix $S$ cannot be indefinite for quantum systems described by Schr\"{o}dinger equation, it can be indefinite for optical systems~\cite{footnote5}.
In the following, we apply the above results to an optical system.

　
\begin{figure*}[]
   \includegraphics[width=18cm]{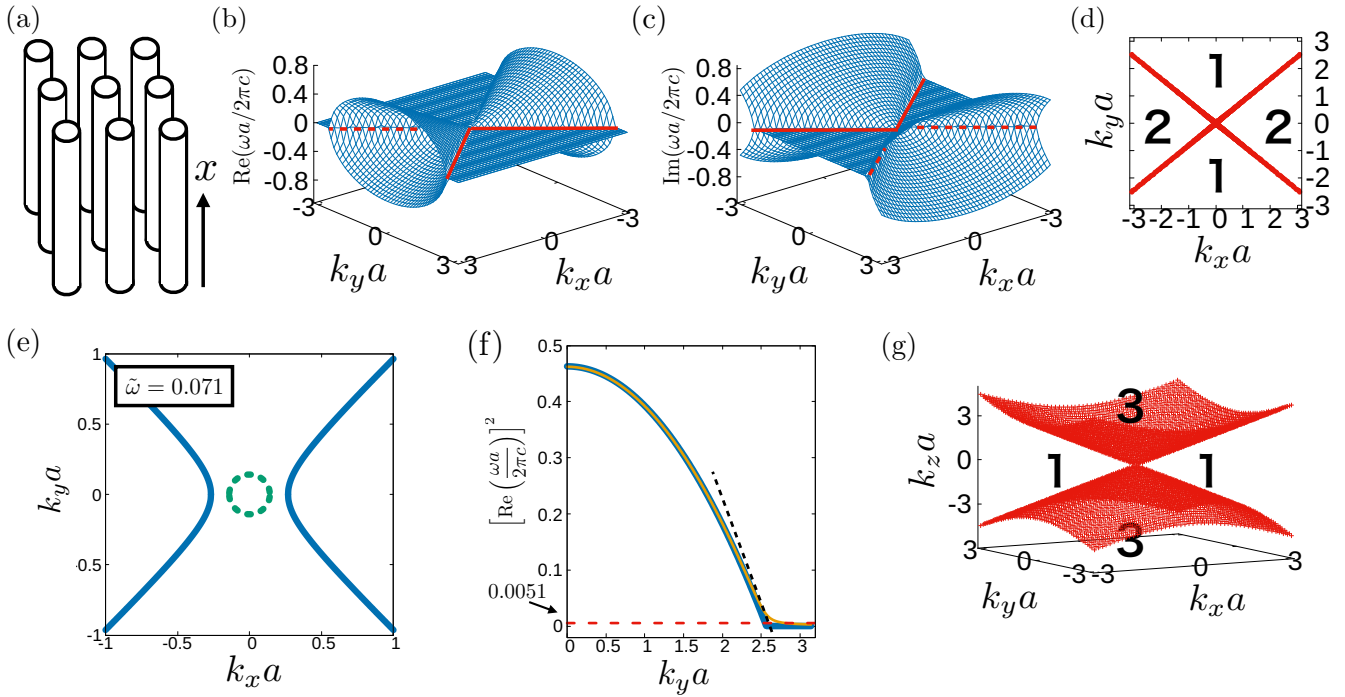}
 \caption{
(a): A sketch of the HMM consist of a metallo-wire structure.
(b) and (c): The real- and imaginary-parts of the dispersion relation of Eq.~(\ref{eq:TEmode}), respectively.
In these panels, red lines indicate the SPERs.
(d): Zero-th Chern number on the $\omega=0$ plane. 
(e): Hyperbolic dispersion with $\tilde{\varepsilon}_{xx}=-0.36$ and $\tilde{\omega}=\omega a/2\pi c=0.071$ (blue solid line) and eliptic dispersion with $\tilde{\varepsilon}_{xx}=0.1$ (green dashed line).
(f):Dispersion relation with square of the dimensionless frequency for $k_x a = \pi$. Data denoted by blue solid line are obtained for $\tilde{\varepsilon}_{xx} = -0.36$, $\tilde{\varepsilon}_{yy}=0.54$, and $\tilde{\mu}_{zz}=1$.
	Data denoted by orange solid line are obtained for $\tilde{\varepsilon}_{xx} = -0.36+i0.016$, $\tilde{\varepsilon}_{yy}=0.54+i0.025$, and $\tilde{\mu}_{zz}=1$.
        We note that the data of the blue line are almost the same as those of the orange line.
	The red dashed line indicates the dimensionless frequency in which the parameter takes the above value.
The black dashed line is a guide to eye.
(g): The Exceptional surface in three-dimensional HMMs with the zero-th Chern number.
 }
 \label{fig:HMM}
\end{figure*}

\vskip\baselineskip
\textit{Application to HMMs.---}
The SPERs described by the GEVP elucidate the origin of the experimentally observed hyperbolic dispersion of the HMMs.   

The HMMs are optical metamaterials with extreme anisotropy. One of the examples is a system of  a metallo-wire structure [Fig.~\ref{fig:HMM}(a)].
In this case,
permittivity for the $x$-direction (permittivity for the $y$- and $z$-direction) satisfies $\epsilon_{xx}<0$ ( $\varepsilon_{yy}>0$ and $\varepsilon_{zz}>0$)~\cite{footnote3}.

In these optical systems, the band structure is described by Maxwell equations; eigenvalues (eigenvectors) correspond to angular frequencies (electromagnetic field).

Here, we consider the two-dimensional system, whose permittivity and permeability are anisotropic.
In this case, EM modes are separable to the TM modes and TE modes. The former is defined as $\bm{E}=(0,0,E_z)^T$ and $\bm{H}=(H_x,H_y,0)^T$ while the latter is defined as $\bm{E}=(E_x,E_y,0)^T$ and $\bm{H}=(0,0,H_z)^T$.
Here, electric (magnetic) filed is denoted by $\bm{E}$ ($\bm{H}$).

Maxwell equations of the TE modes are given by
\begin{align}
\begin{pmatrix}\label{eq:TEmode}
0&-k_y&k_x\\
-k_y&0&0\\
k_x&0&0
\end{pmatrix}
\begin{pmatrix}
H_z\\
E_x\\
E_y
\end{pmatrix}
=\omega
\begin{pmatrix}
\mu_{zz}& & \\
 &\varepsilon_{xx}& \\
 & &\varepsilon_{yy}
 \end{pmatrix}
\begin{pmatrix}
H_z\\
E_x\\
E_y
\end{pmatrix},
\end{align}
with permittivity $\varepsilon$ and permeability $\mu$.
We have assumed that the EM wave is a normal mode which is proportional to $\mathrm{exp}(i\bm{k}\cdot\bm{r}-i\omega t)$.

The eigenvalues of Eq.~(\ref{eq:TEmode}) are given by
 \begin{align}
 \omega_0=0,& & \omega_{\pm}= \pm\sqrt{k_x^{\prime 2}-k_y^{\prime 2}},
 \end{align}
 with $k_x^{\prime}=k_x/\sqrt{\varepsilon_{yy}\mu_{zz}}$ and $k_y^{\prime}=k_y/\sqrt{|\varepsilon_{xx}|\mu_{zz}}$.
 Eigenvectors of each eigenvalue are given by
\begin{align}
 \bm{v}_{\omega_0}=
 \begin{pmatrix}
 0\\
 1\\
 k_y/k_x
 \end{pmatrix},
\ \ \
 \bm{v}_{\omega_{\pm}}=
 \frac{\varepsilon_{yy}}{k_x}
 \begin{pmatrix}
 \omega_{\pm}\\
 k_y/\varepsilon_{xx}\\
 1
 \end{pmatrix}.
\end{align}
We discurd the eigenvector $\bm{v}_{\omega_0}$ because it does not satisfy the Gauss's law.
We also discurd the eigenvector $\bm{v}_{\omega_-}$ in the region of $|k_x^{\prime}|<|k_y^{\prime}|$, in which
$\bm{v}_{\omega_-}$ is amplified because of $\mathrm{Im}(\omega_-)<0$.
In the case of $|k_x^{\prime}|=|k_y^{\prime}|$,
eigenvectors satisfy $\bm{v}_{\omega_0}=\bm{v}_{{\omega_+}}=\bm{v}_{{\omega_-}}$ and all eigenspaces coalesce~\cite{footnote2}.

As a first step to discuss the topological origin of the hyperbolic dispersion, let us show the emergence of the SPERs in this system. 
The emergence of SPERs can be confirmed in the band structure of Eq.~(\ref{eq:TEmode}) [see Fig.~\ref{fig:HMM}(b) and \ref{fig:HMM}(c)]. These data are computed
for $\tilde{\varepsilon}_{xx}=-0.36$, $\tilde{\varepsilon}_{yy}=0.54$, and $\tilde{\mu}_{zz}=1$ which are obtained in Ref.~\cite{W.Ji_PhysRevMaterials(2020)_HMMs}.
$\tilde{\varepsilon}$ and $\tilde{\mu}$ indicate relative permittivity and relative permeability respectively.
At $\omega=0$, in the regions where the condition $|k_x^{\prime}|>|k_y^{\prime}|$ is satisfied,
the eigenvalues become real.
In contrast,
for regions where the condition $|k_y^{\prime}|>|k_x^{\prime}|$ is satisfied,
the eigenvalues become pure imaginary.
The real- and imaginary-parts of the eigenvalues become zero on the $|k_x^{\prime}|=|k_y^{\prime}|$ lines.
This line corresponds to the SPER in the lattice system.
In HMMs, SPERs separate the metallic region and non-metallic region.
These SPERs are robust for perturbation without breaking the Hermiticity of the Maxwell equation because of the topological nature (see below).

The emergence of the above SPERs explains the origin of the hyperbolic dispersion of HMMs.
At $\omega=0$, the isofrequency surface forms the straight lines corresponding to SPERs. 
Increasing $\omega$ from zero,  the structure of the isofrequency surface continuously changes into a hyperbolic structure which is observed in experiments [see blue solid line of Fig.~\ref{fig:HMM}(e)]. 
These results explains that SPERs are the origin of the hyperbolic dispersion of HMMs.
We note that for $\varepsilon_{xx}>0$, $\varepsilon_{yy}>0$, and $\mu_{zz}>0$, the hyperbolic dispersion is not observed [see green dashed line of Fig.~\ref{fig:HMM}(e)] corresponding to the absence of the SPERs.

At $\omega=0$, topological characterization of the metallic region and the non-metallic region of the HMMs is done by computing the zero-th Chern number $N_{0\mathrm{Ch}}$~\cite{T.Yoshida_PRB_2019}.
Figure~\ref{fig:HMM}(c) indicates the zero-th Chern number on the zero-frequency plane.
$\Sigma$ denotes the pseudo-Hermitian operator, and $H_{\Sigma}$ denotes the pseudo-Hermitian matrix (for more details, see Sec.~III of Supplemental Material~\cite{supple}).
The zero-th Chern number takes $N_{0\mathrm{Ch}}=2$ in the non-metallic region and takes $N_{0\mathrm{Ch}}=1$ in the metallic region.

We also expect that the square-root dispersion is one of the experimental signatures of SPERs. 
Here, prior to the experimental observations, we provide theoretical data by approximating that $\varepsilon$ and $\mu$ are independent of $\omega$.
The blue line in Fig.~\ref{fig:HMM}(f) indicates square of dimensionless frequency~\cite{footnote4} for the $\tilde{\varepsilon}_{xx}=-0.36$, $\tilde{\varepsilon}_{yy}=0.54$, and $\tilde{\mu}_{zz}=1$ which are obtained by Ref.~\cite{W.Ji_PhysRevMaterials(2020)_HMMs}.
In Ref.~\cite{W.Ji_PhysRevMaterials(2020)_HMMs}, microwave around 8[GHz] is used.
Near the EP, the dispersion of the square of dimensionless frequency is linear.
Dashed line indicates the dimensionless frequency at which the permittivity and the permeability take the above values.
Therefore, we expect to be able to observe the linear dispersion with the square of the frequency experimentally around the dashed line.
The above band structures exist out of a light cone. 
As the experimental method observe out of the light cone, attenuated total reflection (ATR) method is employed~\cite{otto(1976)_ATR,kretschmann(1968)_ATR,Futamata_ApplOpt(1997)_ATR}.
We note that, for more quantitative prediction, the frequency dependence of $\varepsilon$ and $\mu$ should be taken into account.
Such theoretical works are desired prior to the experimental observation.

Finally, we note that when these HMMs are treated as three-dimensional systems, SPESs form a cone structure in the three-dimensional momentum space.
Figure~\ref{fig:HMM}(g) shows the SPES when the $z$-direction is the anisotropic axis.
These SPESs also explain the characteristic dispersion of three-dimensional HMMs (for details, see Sec.~IV of Supplemental Material~\cite{supple}).

In the above, we have applied the topological band theory to HMMs, which explains the origin of the experimentally observed hyperbolic dispersion of  these systems. 
Decreasing $\omega$ from a finite value to zero, the hyperbolic dispersion asymptotically changes to SPERs.
These SPERs emerging at $\omega=0$, separate metallic region and non-metallic region due to the indefiniteness of the Maxwell equation describing HMMs.

Previous works have analyzed non-Hermitian band structures of HMMs~\cite{Junpeg_PRL_2020_HMMs}. 
We stress, however, that
SPERs and SPESs described by the GEVP reveal the origin of the experimentally observed hyperbolic dispersion of HMMs.

\vskip\baselineskip
　
\textit{Conclusion.---}
In this letter, we have investigated the systems described by the GEVP with Hermitian matrices.
Our analysis has elucidated that non-Hermitian topological band structures emerge in spite of the Hermiticity of the matrices.
For the non-Hermitian topological band structure, the indefiniteness of matrices appearing the left- and right- hand sides is essential.
Remarkably, these SPERs are protected by emergent symmetry; any fine-tuning is not necessary to preserve the symmetry as long as $H$ and $S$ are Hermitian, which is a striking difference of ordinary SPERs.

Furthermore, the above SPERs described by the GEVP reveal the origin of the hyperbolic dispersion of HMMs which is observed in experiments.
In HMMs, SPERs separate the momentum space to the metallic region and the non-metallic region.
Because a hyperbolic isofrequency surface has been observed by angle-resolved reflection spectrum measurements, we also expect that the square-root dispersion of SPERs can be observed which is unique to dispersion of SPERs.

\vskip\baselineskip
\textit{Acknowledgement.---}
We thank Satoshi Iwamoto, Shun Takahashi, and Atsushi Kubo for fruitful discussions.
This work is supported by JSPS
Grant-in-Aid for Scientific
Research on innovative Areas "Discrete
Geometric Analysis for Materials
Design": Grants No.
JP20H04627.
This work is also supported by
JSPS KAKENHI Grants No.
JP17H06138, and No.
JP19K21032.


\begin{thebibliography}{10}
\bibitem{footnote6}
{An example of a system described by the GEVP is a photonic system as we see below.  
Quantum systems are described by the standard eigenvalue problem. For such systems, applying the ordinary topological band theory is more appropriate. 
}
\bibitem{footnote1}
{ When $S$ is definite matrix, $\Sigma$ is the identity matrix, and
  Eq.~(\ref{eq:GEVP}) is transformed to Hermitian standard eigenvalue problem
  by taking $\phi=S^{\prime}\psi, \tilde{H}=S^{\prime -1}HS^{\prime -1}$, and
  $\tilde{H}\phi=E \phi$, where $\tilde{H}$ is Hermitian matrix. Thus, $E$ are
  always real. In the case of $S$ is indefinite and $H$ is definite, we can
  show that $E$ is real by considering the GEVP in the opposite direction.}
\bibitem{footnote9}
{
Consider open quantum systems described by the non-Hermitian eigenvalue problems. 
If the non-Hermitian Hamiltonian preserves the pseudo-Hermiticity [see Eq.~(4)] for such systems, we can mathematically map the non-Hermitian eigenvalue problem to the GEVP.
We note, however, that we mainly focus on photonic systems which are essentially described by the GEVP.
}
\bibitem{footnote5}
{The matrix $S$ of quantum systems are given by $\langle\phi_i|\phi_j\rangle$. This matrix is always positive definite. On the other hand, the matrix $S$ of photonic systems are given by $\langle\phi_i|\varepsilon|\phi_j\rangle$ ($\langle\phi_i|\mu|\phi_j\rangle$), where $\varepsilon$ ($\mu$) is permittivity (permeability). Since $\varepsilon$ ($\mu$) can be negative, the matrix $S$ of photonic systems can be indefinite.}
\bibitem{footnote3}
{ EM waves attenuate with negative $\varepsilon$. This is because electrons in
  the metal screen the electric field. These electrons lose the energy due to
  the resistance in the metal, which results in the imaginary-part of the
  $\varepsilon$. However, our calculation indicates that the small
  imaginary-part of $\varepsilon$ does not significantly change the band
  structure [see Fig.~\ref{fig:HMM}(f)], meaning that the imaginary-part is
  negligible.}
\bibitem{footnote2}
{ In the region of $|k_x^{\prime}|=|k_y^{\prime}|$, $S^{-1}H$ is written as
  \begin{equation*} S^{-1}H(k_x^{\prime}=k_y^{\prime})= \begin{pmatrix} 0&1&0\\
  0&0&1\\ 0&0&0 \end{pmatrix}, \end{equation*} which is non-diagonalizible.}
\bibitem{footnote4}
{ We note that the square-root dispersion survives even when the imaginary part
  of $\varepsilon$ and $\mu$ are taken into account [see the orange line of
  Fig.~\ref{fig:HMM}(f)].}
\end{thebibliography}

\clearpage

\renewcommand{\thesection}{S\arabic{section}}
\renewcommand{\theequation}{S\arabic{equation}}
\setcounter{equation}{0}
\renewcommand{\thefigure}{S\arabic{figure}}
\setcounter{figure}{0}
\renewcommand{\thetable}{S\arabic{table}}
\setcounter{table}{0}
\makeatletter
\c@secnumdepth = 2
\makeatother

\onecolumngrid
\begin{center}
 {\large \textmd{Supplemental Materials:} \\[0.3em]
 {\bfseries Topological band theory of a generalized eigenvalue problem with Hermitian matrices: Symmetry-protected exceptional rings with emergent symmetry}}
\end{center}
\setcounter{page}{1}

\section{Complex conjugate eigenvalues of pseudo-Hermitian matrix}
\label{sec: pseudo-H app}

In this section, we show that the complex eigenvalues of the pseudo-Hermitian matrix appear as complex conjugate pairs.
The condition of pseudo-Hermite is defined as
\begin{equation}\label{eq:pseudo-Hermite}
\Sigma^{-1}H^{\dagger}\Sigma=H,
\end{equation}
with operator $\Sigma$.
Here, let us consider the eigenvalue problem
\begin{align}\label{eq:righteigeq}
H|R_n\rangle=E_n|R_n\rangle,\\\notag
\langle L_n|H=E_n\langle L_n|,
\end{align}
where $|R_n\rangle$ ($\langle L_n|$) are right (left) eigenstates, and $E_n$ are eigenvalues. $n$ indicate the label of the eigenvalues.
From Eq.~(\ref{eq:pseudo-Hermite}) and Eq.~(\ref{eq:righteigeq}), we can be also obtained eigenvalue problem with eigenstates $\Sigma|L_n\rangle$,
\begin{equation}\label{eq:eigeq2}
H(\Sigma|L_n\rangle)=\Sigma H^{\dagger}|L_n\rangle=\Sigma(\langle L_n|H)^{\dagger}=E^*_n(\Sigma|L_n\rangle),
\end{equation}
From Eq.~(\ref{eq:eigeq2}), we can see the fact that $\Sigma$ maps the eigenstate of eigenvalue $E_n$ to the eigenstate of eigenvalue $E_n^*$.
Therefore, complex eigenvalues of the pseudo-Hermitian matrix are given by complex conjugate pairs.

\section{Eigenvalue of the two-band model}
\label{sec: 2-band model app}

In the main text, we have introduced the $2\times2$ model of a GEVP with diagonal overlap matrix $S$.
In this section, we consider the $2\times2$ model with the off-diagonal term of $S$,
\begin{equation}
\begin{pmatrix}
\epsilon+m_L&vf_{\bm{k}}\\
vf_{\bm{k}}^* & \epsilon-m_L
\end{pmatrix}
\psi=E
\begin{pmatrix}
1+m_R&wf_{\bm{k}}\\
wf_{\bm{k}}^* & 1-m_R
\end{pmatrix}
\psi,
\end{equation}
with the onsite potential $\epsilon$ and hopping $v$ and $w$.
The difference of the  onsite potentials are parametrized by $m_L$ and $m_R$.
The eigenvalue (eigenvector) are denoted by $E (\psi)$.
where $E$ is eigenvalue, $\psi$ is eigenvector, $\epsilon$ is onsite term, $v$ and $w$ are hopping term of left and right matrix respectively, $m_L$ and $m_R$ describe the potential defference of red site and blue site (Fig.~1(a)),
and $f_{\bm{k}}$ is defined as $f_{\bm{k}}=(1+e^{i\bm{k}\cdot\bm{t_1}+e^{\bm{k}\cdot\bm{t_2}}})$.

Let us discuss the band structure.
First, we take left side matrix $H$, right side matrix $S$ and diagonalize matrix $S$ by the unitary matrix $U_S$,
\begin{equation}
U_S^{-1}H U_S\psi =E U_S^{-1} S U_S\psi.
\end{equation}
$U_S^{-1}H U_S$ and $U_S^{-1} S U_S$ are taken $\tilde{H}, \tilde{S}$ to simplify the symbols.
Here, eigenvalues of matrix $S$ are given as $s_\pm=1\pm \sqrt{|wf(\bm{k})|^2+m_R^2}$.

Next, matrix $S$ is decomposed as,
\begin{align}
\tilde{S}&=\tilde{S}^{\frac{1}{2}}\Sigma \tilde{S}^{\frac{1}{2}}
=
\begin{pmatrix}
\sqrt{|s_+|}&0\\
0&\sqrt{|s_-|}\\
\end{pmatrix}
\Sigma
\begin{pmatrix}
\sqrt{|s_+|}&0\\
0&\sqrt{|s_-|}\\
\end{pmatrix}.
\end{align}
where $S^{\frac{1}{2}}$ and $\Sigma$ consist of the square root of the absolute value of eigenvalue and sign of eigenvalue of matrix $S$.
For $\mathrm{det}S>0$, matrix $\Sigma$ becomes identity matrix. 
In this case, a GEVP transformed Hermitian standard eigenvalue problem as follows,
\begin{equation}
\left(\tilde{S}^{-\frac{1}{2}}\tilde{H} \tilde{S}^{-\frac{1}{2}}\right) \left(\tilde{S}^{\frac{1}{2}}\psi\right)=E\left(\tilde{S}^{\frac{1}{2}}\psi\right).
\end{equation}
where matrix $\left(\tilde{S}^{-\frac{1}{2}}\tilde{H} \tilde{S}^{-\frac{1}{2}}\right) \left(\tilde{S}^{\frac{1}{2}}\right)$ is 
\begin{align}
\tilde{S}^{-\frac{1}{2}}\tilde{H} \tilde{S}^{-\frac{1}{2}}=
\begin{pmatrix}
\frac{\epsilon+m_L}{|s_+|}&\frac{vf_{\bm{k}}}{\sqrt{|s_+|}\sqrt{|s_-|}}\\
\frac{vf_{\bm{k}}}{\sqrt{|s_+|}\sqrt{|s_-|}}&\frac{\epsilon-m_L}{|s_-|}
\end{pmatrix}.
\end{align}
Eigenvalues are given by
\begin{align}\label{eigval-def}
E&=\frac{1}{2}\left[\left(\frac{\epsilon+m_L}{|s_+|}+\frac{\epsilon-m_L}{|s_-|}\right)
\pm
\sqrt{\left(\frac{\epsilon+m_L}{|s_+|}-\frac{\epsilon-m_L}{|s_-|}\right)^2+\frac{4v^2|f_{\bm{k}}|^2}{|s_+||s_-|}}\right].
\end{align}
However,
For $\mathrm{det}S<0$, matrix $\Sigma$ becomes 
\begin{equation}
\Sigma=\sigma_3=
\begin{pmatrix}
1&0\\
0&-1
\end{pmatrix}.
\end{equation}
 In this case, the GEVP cannot be transformed in to a Hermitian standard eigenvalue problem but into a non-Hermitian standard eigenvalue problem as follows,
$\sigma_3\left(\tilde{S}^{-\frac{1}{2}}\tilde{H} \tilde{S}^{-\frac{1}{2}}\right) \left(\tilde{S}^{\frac{1}{2}}\psi\right)=E\left(\tilde{S}^{\frac{1}{2}}\psi\right).$
We can easily prove the matrix $\tilde{S}^{-\frac{1}{2}}\tilde{H} \tilde{S}^{-\frac{1}{2}}$ satisfy the pseudo-Hermite condition,
\begin{equation}
\sigma_3\left(\sigma_3\tilde{S}^{-\frac{1}{2}}\tilde{H} \tilde{S}^{-\frac{1}{2}}\right)\sigma_3
=
\left(\tilde{S}^{-\frac{1}{2}}\tilde{H} \tilde{S}^{-\frac{1}{2}}\sigma_3\right)
=
\left(\sigma_3\tilde{S}^{-\frac{1}{2}}\tilde{H} \tilde{S}^{-\frac{1}{2}}\right)^{\dagger},
\end{equation}
where we used Hermiticity of matrix $\tilde{H}$, $\tilde{S}^{\frac{1}{2}}$, and $\sigma_3$.

The eigenvalues of this model are described as
\begin{align}\label{eigval-indef}
E&=\frac{1}{2}\left[\left(\frac{\epsilon+m_L}{|s_+|}-\frac{\epsilon-m_L}{|s_-|}\right)
\pm
\sqrt{\left(\frac{\epsilon+m_L}{|s_+|}+\frac{\epsilon-m_L}{|s_-|}\right)^2-\frac{4v^2|f_{\bm{k}}|^2}{|s_+||s_-|}}\right].
\end{align}
Here, eigenvalue (\ref{eigval-indef}) can be complex though eigenvalue (\ref{eigval-def}) cannot be complex [see Fig.~1].

\begin{figure}[h]
   \includegraphics[width=15cm]{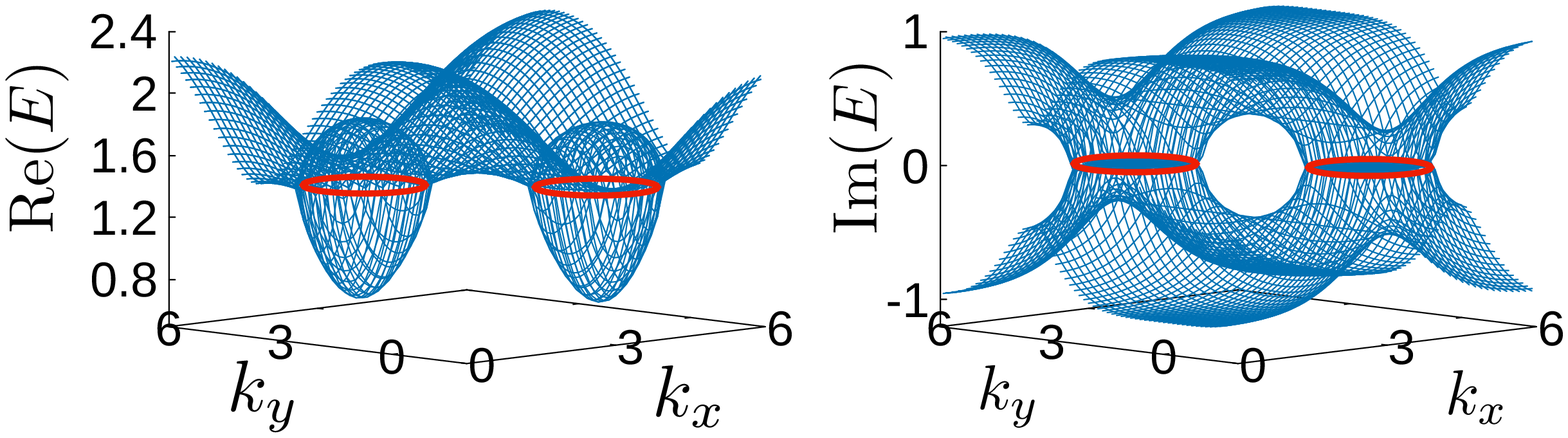}
 \caption{
	 Band structure of the $2\times2$ model with the off-diagonal term of $S$. Parameters are chosed in $\epsilon=2.0$, $v=1.0$, $m_L=2.2$, $w=0.3$, and $m_R=1.3$. Red lines indicate the SPERs. These figures demonstrate the robustness of SPERs described by the GEVP.}
 \label{fig:SPER2}
\end{figure}


\section{Computation of the zeroth-Chern number}
\label{sec: 0thCh_HMM_app}
In this section, we compute the zeroth-Chern number of HMMs.
Here, we consider the case of TE modes with $\mu_{zz}=1$, $\varepsilon_{xx}=-1$, and $\varepsilon_{yy}=1$.
The matrix of left-hand (right-hand) $H$ ($\Sigma$) is given by
\begin{align}
H=
\begin{pmatrix}
0&-k_y&k_x\\
-k_y&0&0\\
k_x&0&0
\end{pmatrix}, 
\ \ \
\Sigma=
\begin{pmatrix}
1&0&0\\
0&-1&0\\
0&0&1
\end{pmatrix}.
\end{align}
First, we decompose $H_{\Sigma}=\Sigma H$ as follows,
\begin{align}
H_{\Sigma}=\hat{R}\hat{E}\hat{L}^{\dagger}=
\begin{pmatrix}
\bm{v}_{R,0},&\bm{v}_{R,\omega_+},&\bm{v}_{R,\omega_-}
\end{pmatrix}
\begin{pmatrix}
\omega_0&0&0\\
0&\omega_+&0\\
0&0&\omega_-
\end{pmatrix}
\begin{pmatrix}
\bm{v}_{L,\omega_0}^{\dagger}\\
\bm{v}_{L,\omega_+}^{\dagger}\\
\bm{v}_{L,\omega_-}^{\dagger}
\end{pmatrix},
\end{align}
with $\omega_0=0$, and $\omega_{\pm}=\pm\sqrt{k_y^2-k_x^2}$.
On the SPERs, the condition of $\mathrm{det}(\Sigma H_{\Sigma})=0$ is satisfied.
However, our Hamiltonian is always $\mathrm{det}(\Sigma H_{\Sigma})=0$ due to the existence of $\omega_0$.
In order to characterize  the SPERs, we define the modified Hamiltonian $\tilde{H}_\Sigma$, which $\omega_0$ is shifted by $\delta$,
\begin{align}
\tilde{H}_{\Sigma}=\hat{R}\hat{\tilde{E}}\hat{L}^{\dagger}=
\begin{pmatrix}
\bm{v}_{R,0},&\bm{v}_{R,\omega_+},&\bm{v}_{R,\omega_-}
\end{pmatrix}
\begin{pmatrix}
\omega_0+\delta&0&0\\
0&\omega_+&0\\
0&0&\omega_-
\end{pmatrix}
\begin{pmatrix}
\bm{v}_{L,\omega_0}^{\dagger}\\
\bm{v}_{L,\omega_+}^{\dagger}\\
\bm{v}_{L,\omega_-}^{\dagger}
\end{pmatrix}.
\end{align}
Therefore, Hermitian matrix $\Sigma\tilde{H}_{\Sigma}$ is given by
\begin{align}
\Sigma\tilde{H}_{\Sigma}=
\begin{pmatrix}
0&-k_y&k_x\\
-k_y&-\frac{k_x^2}{k_x^2-k_y^2}\delta&\frac{k_xk_y}{k_x^2-k_y^2}\delta\\
k_x&\frac{k_xk_y}{k_x^2-k_y^2}\delta&-\frac{k_y^2}{k_x^2-k_y^2}\delta
\end{pmatrix}.
\end{align}
In the main text, we characterized the SPERs by the zeroth-Chern number of $\Sigma\tilde{H}_{\Sigma}$ which is the number of negative eigenvalues.
We note that the zeroth-Chern number in Figs.~2(b) and 2(e) are computed for $\delta>0$.
In the case for $\delta<0$, the region of $N_{0\mathrm{Ch}}=1$ and that of $N_{0\mathrm{Ch}}=2$ are fliped.

\section{SPES and three-dimensional hyperbolic dispersion}
\label{sec: SPES_app}
\begin{figure}[h]
   \includegraphics[width=18cm]{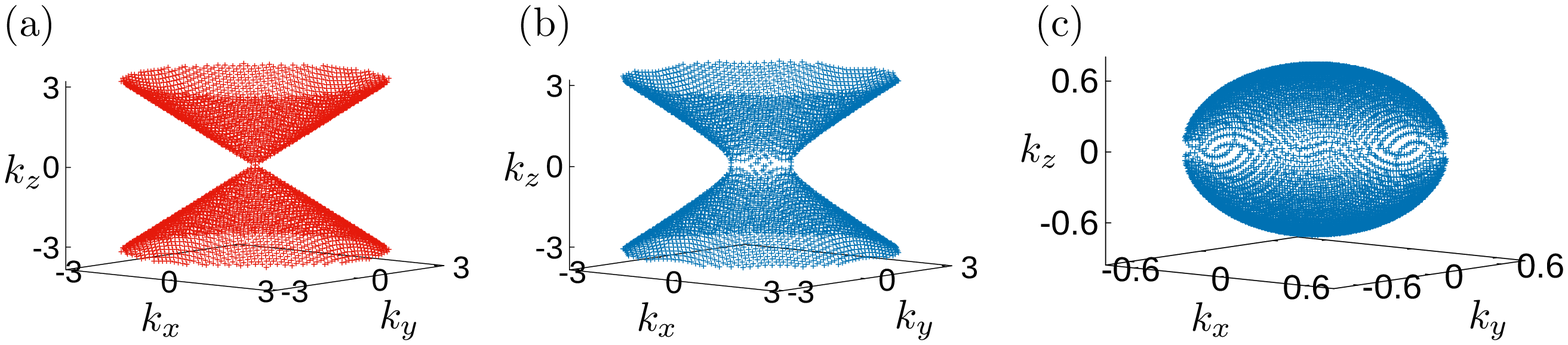}
 \caption{
	(a): The SPES on the $\omega=0$ surface in the three-dimensional momentum space.
	(b): Hyperbolic dispersion in three-dimensional momentum space with $\omega=0.5$.
	(c): Eliptic dispersion with $\varepsilon_{xx}>0$. 
 }
 \label{fig:SPES}
\end{figure}

In the main text, we have shown that SPERs  at $\omega=0$ are the origin of the hyperbolic dispersion.
In this section, we show that SPESs at $\omega=0$ are the origin of hyperbolic dispersion in the three-dimensional momentum space.
Here, we analyze three-dimensional Maxwell equations,
\begin{equation}\label{eq:Maxwell6x6}
\begin{pmatrix}
0&\bm{k}\times\\
-\bm{k}\times&0
\end{pmatrix}
\begin{pmatrix}
\bm{E}\\
\bm{H}
\end{pmatrix}
=\omega
\begin{pmatrix}
\varepsilon&0\\
0&\mu
\end{pmatrix}
\begin{pmatrix}
\bm{E}\\
\bm{H}
\end{pmatrix}. 
\end{equation}
We assume that $z$-direction is anisotropic axis. 
The permittivity and the permeability are given by
\begin{align}
\varepsilon=
\begin{pmatrix}
1&0&0\\
0&1&0\\
0&0&-1
\end{pmatrix},
 \ \ \
 \mu=
 \begin{pmatrix}
 1&0&0\\
0&1&0\\
0&0&1
\end{pmatrix}.
\end{align}
In this case, non-zero eigenvalues are given by
\begin{equation}
\omega=\pm\sqrt{k_x^2+k_y^2-k_z^2}.
\end{equation}
A surface satisfying $k_x^2+k_y^2=k_z^2$ corresponds to a SPES [see Fig.~\ref{fig:SPES}(a)].
Solving the above equation with respect to $k_z$ we obtain
\begin{equation}
k_z=\pm\sqrt{k_x^2+k_y^2-\omega^2}.
\end{equation}
When SPESs emerge at $\omega=0$, hyperbolic dispersion in three-dimensional momentum space emerges for $\omega\ne 0$ [see Fig.~\ref{fig:SPES}(b)].
For $\varepsilon=\begin{pmatrix}1&0&0\\0&1&0\\0&0&1\end{pmatrix}$, the matrix in the right-hand side of Eq.~(\ref{eq:Maxwell6x6}) is positive definite. 
In this case, the SPES vanishes.
Correspondingly, hyperbolic dispersion in three-dimensional momentum space also vanishs [see Fig.~\ref{fig:SPES}(c)].

　
　
　

\end{document}